# Knowledge Management Strategies and Emerging Technologies - an Overview of the Underpinning Concepts

*Siddhartha Paul Tiwari*, Google Asia Pacific, Mapletree Business City, Singapore

**Abstract**

Among the essential elements of knowledge management is the use of information and data, as well as the knowledge, skills, and abilities inherent within communities, as well as their ideas, commitments, and motivations for making good decisions as emerging technologies become more prevalent. Numerous leading social scientists in this field have asserted that organizational knowledge should be regarded as a strategic asset. There is a growing awareness of the importance of gathering, locating, capturing, and sharing collective knowledge and expertise of societies, and societies are urged to develop effective and efficient methods of gathering, locating, capturing, and sharing that knowledge in order to deal with problems and to benefit from opportunities. People living in many countries and regions are interested in implementing knowledge management processes and technologies, and many of them have included knowledge management as an integral part of their overall development strategies. The management of knowledge plays an increasingly important role in global economic development (Bell, 1973, 1978). In order to remain relevant in the modern world, organizations should not ignore knowledge management and emerging technologies.

1. **Introduction.** Traditionally, knowledge is believed to be distinct from data and information. As technology advances, new possibilities are becoming available. The integration and sharing of highly distributed information is a prerequisite for achieving effective performance and growth in knowledge-intensive societies. Over the past few decades, as the society has grown and evolved in sync with evolving technology, knowledge management has become one of the most important aspects enabling the productivity and competitiveness of society and organizations to increase to the same level regardless of market conditions. According to McKern (1996), powerful forces lead to powerful outcomes which are shaping the economic and business climate in the world and many people are calling for a fundamental change in the way we think and act about the way we run our organizations. Taking into account the benefits of emerging technologies such as the creation of reusable services that allow organizations to consume these services on their own, knowledge management is also considered to be a significant element.

2. **Knowledge Management and Emerging Technologies.** Throughout an organization there is a collection of knowledge, which is created and maintained by the members, and with the emergence of new technologies, a greater amount of knowledge is being produced. When a knowledge management strategy is implemented throughout the organization, in addition to improving service levels, it will also increase customer satisfaction and efficiency. To fully comprehend emerging technologies, we must first determine what they are, what they do, and what they mean for the organization. The term emerging technology refers to technologies which, in terms of their development, practical application, or both, have not yet reached their full potential, as such they are emerging figuratively into prominence from obscurity or nonexistence, for example artificial intelligence. Emerging technology can refer to new technical developments or to the continuous improvement of an existing technology. The future of artificial intelligence (AI) and knowledge management has been discussed at length, as well as the relation between the two (Liebowitz, 2001). Knowledge ManagementIn a variety of situations, knowledge management is the fastest and easiest way for an organization to add value.

Knowledge management is also becoming increasingly important due to the rapid changes of needs, processes, and the process of bringing management systems into dialogue when the proposed strategy is aligned with the emerging technologies employed by the organization. Having innovative capabilities arising from emerging technologies, which are meeting the needs and expectations of society, allows organizations to remain competitive, consistent in meeting expectations and needs, and differentiate systems based on conceptualized knowledge. The organization can use this knowledge to more effectively communicate with its partners and customers, which in turn allows them to make better decisions in response to an issue when addressing it. Emerging technologies have made it possible for partners to access the information gathered from various sources, which is helpful for the decision-making process.

3. **Emerging Technologies Underpinning Knowledge Management.** Since the 1990s, technology has played a major role in transforming almost everything about organizations' day-to-day operations. Increasingly, organizations are introducing knowledge management initiatives in order to spread their programs. Davenport (1998) states that knowledge is framed experiences, values, and context as expert insights that provide a framework for evaluating & incorporating new information. In order for the organizations to adopt a knowledge management approach in line with the emerging technologies, the first fundamental aspect is an increased level of responsiveness to the organization's customers and partners. In terms of the development and expansion of a business, the knowledge that is gathered about the people within the organization is valuable for the organization. Obtaining this knowledge may be difficult in some instances, unless emerging technologies make it more straightforward to collect and assemble this information from a variety of sources. It is possible, for instance, to collect different applications and data structures and assemble conclusive demands, and then develop definitive requests based on those demands.

4. **Challenges Associated with Knowledge Management and Emerging Technologies.** There are a number of dynamic challenges that have emerged from the emergence of new technologies that need to be addressed if knowledge management processes are to become more effective for organizations. It is crucial that knowledge management strategists within organizations consider these challenges and develop strategies that ensure that the recommendations they make in the area of knowledge management help them to meet their obligations towards consumers and other stakeholders. It is important to note here that there is an issue that needs to be addressed. Furthermore, it is also important to keep in mind that one of the challenges we are likely to face when we implement knowledge management concerning emerging technologies and connected systems has to do with the fact that we are dealing with relatively new systems. In addition to those reasons, it is also recommended that we keep in mind that even though there are standards governing how these systems should be implemented, they are still a bit underdeveloped. There is a possibility that in the future, an organization's knowledge strategy may be faced with ethical dilemmas arising from the use of new technologies associated with its knowledge strategy. In summary, knowledge management is evolving from a tool for process improvements and marginal efficiency to a key element of innovation that will underpin the future of organizations worldwide.

5. **Solution.** Whenever it comes to knowledge management, emerging technologies should suggest options to users, rather than forcing them on them. Knowledge management systems should not be controlled by technology, neither should technology produce direct effects, nor should technology dictate how knowledge management systems should be managed. There is no question that at the end of the day, it is ultimately up to society to determine what they want to do with emerging technologies that impact knowledge management. What knowledge management technology should be able to do is simply shed a light on the numerous possibilities that are available to them regarding knowledge creation, knowledge exchange, and knowledge management. It is ultimately the decision of intelligent individuals handling these complex systems that will ultimately determine the outcome. It is no secret that emerging technologies are a necessary part of creating a knowledge strategy for organizations, as is the use of knowledge systems. Nonetheless, the purpose of the knowledge management systems and tools is ultimately to assist the organization in achieving a competitive advantage and shouldn't be used to control or dictate the knowledge management pipeline. There are many technical aspects to the new technology that are causing it to be highly fragmented, with a range of uses in various parts of society. Consequently, there is no one solution that will fit all organizations utilizing new technologies for knowledge management and we need to look at each case on its own merit, which may serve as the basis of future research for the societies.

6. **Conclusion.** In the context of emerging technologies, the function of knowledge management is to improve responsiveness to partners and customers, increase the ability to solve problems, and develop individual competencies. Knowledge management is one of the greatest assets of the organization; it is able to make well-informed decisions and to retain the knowledge in the organization for future use if needed. Incorporating a knowledge management system and emerging technologies within an organization gives insight into the opportunities that may arise, which prompts responses to selected issues and services.

   Increasing the accuracy of decision-making through the use of knowledge management and emerging technologies is crucial for an organization's success. To succeed in this process, data must be collected, the first step in the process. Having collected and organized the data, and having organized the results in an appropriate manner, it is important to draw conclusions about the implications of the collected data. Within a few years, the majority of knowledge will likely come from new technologies and computerized systems, enabling organizations to make more informed decisions and to manage information more efficiently and effectively.